\documentstyle[manuscript,aps]{revtex}
\begin{document}

\title{Comment on ``Non-thermalizability of a Quantum Field Theory''}

\author{S. Deser$^{(a)}$,  G. Dunne$^{(b)}$, L. Griguolo$^{(c)}$,
       K. Lee$^{(d)}$, C. Lu$^{(d)}$ and D. Seminara$^{(a)}$}
\address{\it $^{(a)}$ Department of  Physics, Brandeis University,
                        Waltham, MA 02254\\
             $^{(b)}$ Department of Physics, University of Connecticut,
                         Storrs, CT 06269\\
	 $^{(c)}$ Center for Theoretical Physics, Massachusetts Institute of
Technology, Cambridge, MA 02139\\
         $^{(d)}$ Department of Physics, Columbia University,
                         New York, NY 10027}
\maketitle

\begin{abstract}
We point out that the claims made in the paper ``Non-thermalizability of a
 Quantum Field Theory'' (hep-th/9802008) by C. R. Hagen are irrelevant to our
 recent results concerning large gauge invariance of the effective action in
thermal QED.
\end{abstract}

\vskip 1cm



There has been much recent interest in finite temperature
 Chern-Simons theories
 \cite{dunne1,deser1,schaposnik1,aitchison1,gonzalez,jim,jackiw},
 focussing on an inherent incompatibility between large gauge
 invariance and conventional finite temperature perturbation theory. For
 example, thermal perturbation theory leads to induced Chern-Simons
 terms with temperature dependent coefficients, a result that, on the face of
 it, seems to imply a violation of large gauge invariance. However, as first
 shown in \cite{dunne1} in the context of an exactly solvable $0+1$ dimensional
 Chern-Simons model, the finite temperature effective action contains new
 induced
 terms, other than the Chern-Simons term, each of which is temperature
 dependent. When these are resummed, the full effective action satisfies large
 gauge invariance, despite the fact that large gauge invariance is violated at
 any finite order in perturbation theory. An elegant way to
 understand this incompatibility between large gauge invariance and finite
 temperature perturbation theory is through zeta function regularization
 \cite{deser1}. These results may be extended to $2+1$ dimensional
 thermal QED in the presence of particular background gauge fields which
 support large gauge transformations
 \cite{deser1,schaposnik1,aitchison1,gonzalez}.

The recent paper of Hagen \cite{hagen} attempts a full solution of the $0+1$
 charged fermion-Chern Simons model, and then claims that the aforementioned
 results for the  $0+1$
 dimensional model of \cite{dunne1} (and implicitly therefore also for
 \cite{deser1,schaposnik1,aitchison1,gonzalez}) are incorrect. However, this
 disagreement is due to a simple but fundamental error of interpretation.
 Whatever the validity of the calculation in \cite{hagen}, it is totally
 irrelevant to the subject of
 \cite{dunne1,deser1,schaposnik1,aitchison1,gonzalez,jim}, namely the
 properties of the effective gauge field action obtained by integrating out the
 charged sources: by the very definition of effective actions, the gauge field
 there is of course NOT to be evaluated on any a priori mass shell!

\end{document}